\documentclass[fleqn,12pt,twoside]{article}
\usepackage[headings]{espcrc1}

\usepackage{graphicx}
\usepackage[figuresright]{rotating}

\usepackage{amsmath}
\usepackage{verbatim}

\def\slashed#1{\kern+0.1em /\kern-0.55em #1}

\title{Testing Algorithms for Finite Temperature Lattice QCD}

\author{M. Cheng\address[Columbia]{Department of Physics, Columbia University, New York, NY 10027}, M. A. Clark\address[BU]{Center for Computational Science, Boston University, Boston, MA 02215}, C. Jung\addressmark[Columbia]\address[BNL]{Physics Department, Brookhaven National Laboratory, Upton, NY 11973}, R. D. Mawhinney\addressmark[Columbia]}

\begin{document}

\maketitle

\begin{abstract}
We discuss recent algorithmic improvements in simulating finite temperature QCD on a lattice.  In particular, the Rational Hybrid Monte Carlo(RHMC) algorithm is employed to generate lattice configurations for 2+1 flavor QCD.  Unlike the Hybrid R Algorithm, RHMC is reversible, admitting a Metropolis accept/reject step that eliminates the $\mathcal{O}(\delta t^2)$ errors inherent in the R Algorithm.  We also employ several algorithmic speed-ups, including multiple time scales, the use of a more efficient numerical integrator, and Hasenbusch pre-conditioning of the fermion force.
\end{abstract}

\section{Finite Temperature Lattice QCD}

Lattice QCD can be used to probe the non-perturbative properties of QCD at finite temperature.  The partition function, $Z(V,T)$ can be written as:
\begin{equation}
\mathcal{Z} (V,T) =
\int \;{\cal D} A_\nu {\cal D}\bar{\psi}{\cal D}\psi\;
\mbox{exp}(-S_E(V,T)) \quad
\label{partZ}
\end{equation}
This expression corresponds to the path integral for a four-dimensional field theory with temporal extent $\tau = 1/T$.

The RBC-Bielefeld Collaboration is currently performing large-scale simulations of finite temperature lattice QCD to calculate important physical quantities such as the transition temperature $T_c$ and the QCD Equation of State (EoS).  Here, we discuss the algorithmic improvements used in these simulations.

\section{Hybrid Molecular Dynamics (HMD)}
Hybrid Molecular dynamics (HMD) uses both a molecular dynamics evolution and Monte Carlo techniques to sample configuration space.  Start by introducing a field of traceless, Hermitian SU(3) matrices, $P_i$.  The $P_i$ act like conjugate momenta to the gauge links $U_i$, so the pseudo-Hamiltonian is: 
\begin{equation}
\mathcal{H} = \sum_i Tr(P_i^2/2) + S_{QCD}
\end{equation}
Hamilton's equations are used to evolve the system along an (approximately) energy-conserving trajectory of fixed length ($\tau $= 0.5 or 1.0).  At the end of a trajectory, $P_i$ are randomly refreshed by coupling to a heat bath, sending the next trajectory in a different direction.  The amount of phase space sampled $\sim N$, compared to $\sim \sqrt{N}$ for random-walk type algorithms.

In our case, using staggered fermions, we can integrate out the Grassmann fields, which gives ($n_f$ denoting the number of degenerate quark flavors):
\begin{equation}
\mathcal{Z} = \int [\mathcal{D}U]\;\mbox{det}(\mathcal{M})^{n_f/4}\;\mbox{exp}(-S_g)
\end{equation}
For $n_f=4$, one employs the $\Phi$ algorithm\cite{Gottlieb:1987mq}, which uses pseudo-fermion fields to evaluate the determinant of the fermion matrix.  Since the $\Phi$ algorithm satisfies detailed balance, we can use a reversible integrator that admits a Metropolis accept/reject step.  

This is not the case for $n_f \neq 4$, where the fermion determinant results in a non-local term in the effective action.  The R algorithm\cite{Gottlieb:1987mq} evaluates this non-local term using a noisy estimator.  However, this introduces unacceptably large $\mathcal{O}(\delta t)$ discretization errors when numerically integrating the equations of motion.  These errors can be reduced to $\mathcal{O}(\delta t^2)$, but only at the cost of using a non-reversible numerical integrator.

\section{Rational Hybrid Monte Carlo (RHMC)}
Recently, it was realized that one can use an optimal, rational approximation to evaluate $\mbox{det}(\mathcal{M})^{n_f/4}$\cite{Kennedy:1998cu}.  The idea is to use a rational approximation that is valid to some arbitrary precision over the spectral range of the fermion operator.
\begin{equation}
\mbox{det}(\mathcal{M})^{n_f/4} = \int[\mathcal{D}\bar{\phi}][\mathcal{D}\phi]\mbox{exp}(\bar{\phi}\mathcal{M}^{-n_f/4}\phi) = \int[\mathcal{D}\bar{\phi}][\mathcal{D}\phi]\mbox{exp}(\bar{\phi}r^2(\mathcal{M})\phi)
\end{equation}
where $r(x) \approx x^{-n_f/8}$.  The rational approximation allows the use of a reversible integrator with a Metropolis accept/reject step, regardless of the number of flavors\cite{Clark:2003na}.  RHMC is exact - it lacks the discretization errors associated with the R Algorithm.  The RHMC is also more efficient - one can use a much larger integration step size than the one typically used for the R Algorithm ($\delta t \approx 0.4 m_f$).  As seen in Fig. 1, recent comparisons of the two algorithms show that step-size errors can sometimes be quite severe, especially for finite temperature simulations.\cite{Clark:2005sq,deForcrand:2006pv}

\begin{figure}[tb]
\begin{minipage}[c]{0.47\textwidth}
\includegraphics[width=\textwidth]{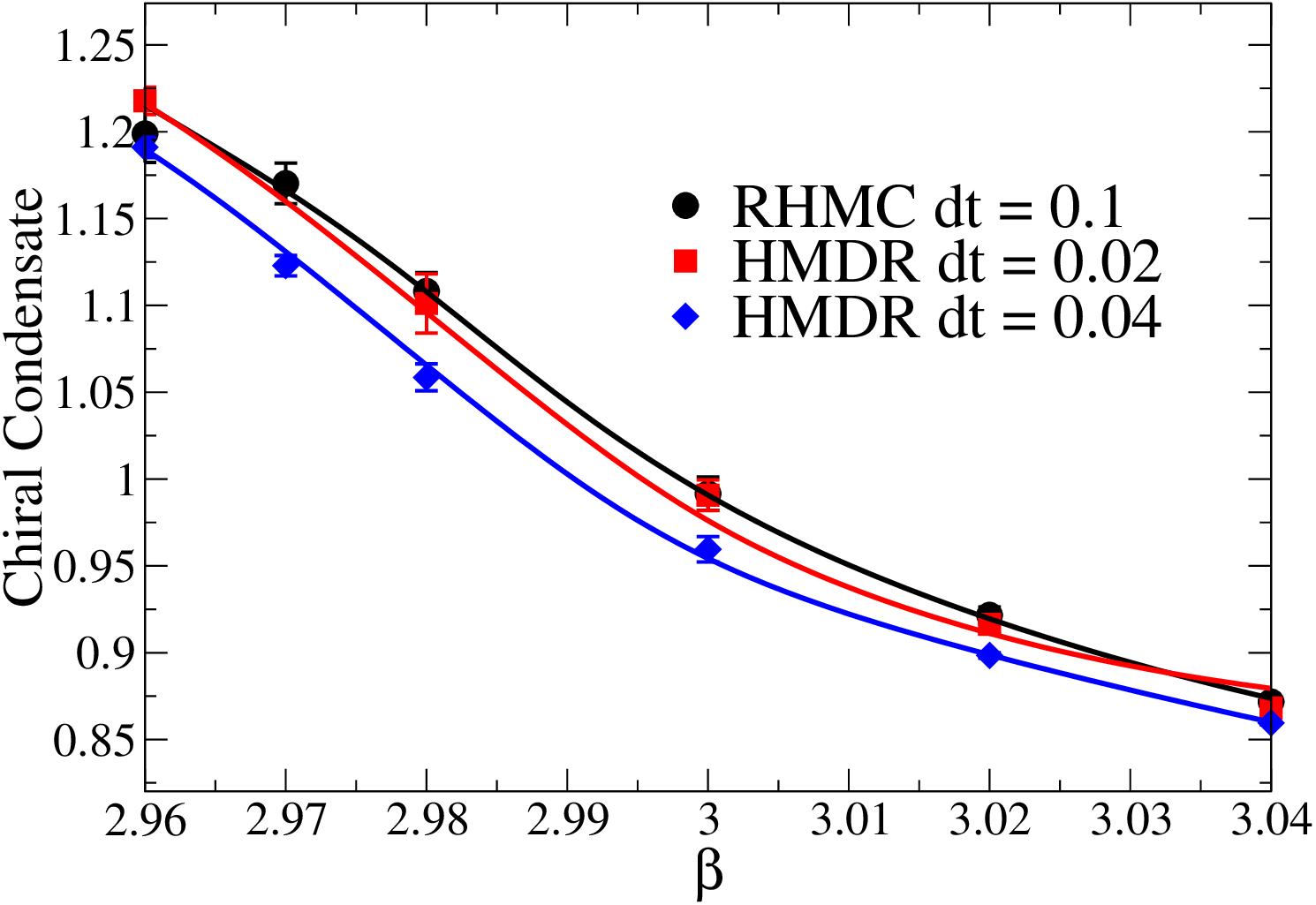}
\label{ralg}
\vspace{-1.2cm}
\caption{Finite step size errors in the chiral condensate using the R Algorithm. 3f QCD, $8^3\times 4$ volume, p4fat7 fermion action, $m_q a = 0.1$}
\end{minipage}
\hspace{\fill}
\begin{minipage}[c]{0.47\textwidth}
\vspace{0.2cm}
\includegraphics[width=\textwidth]{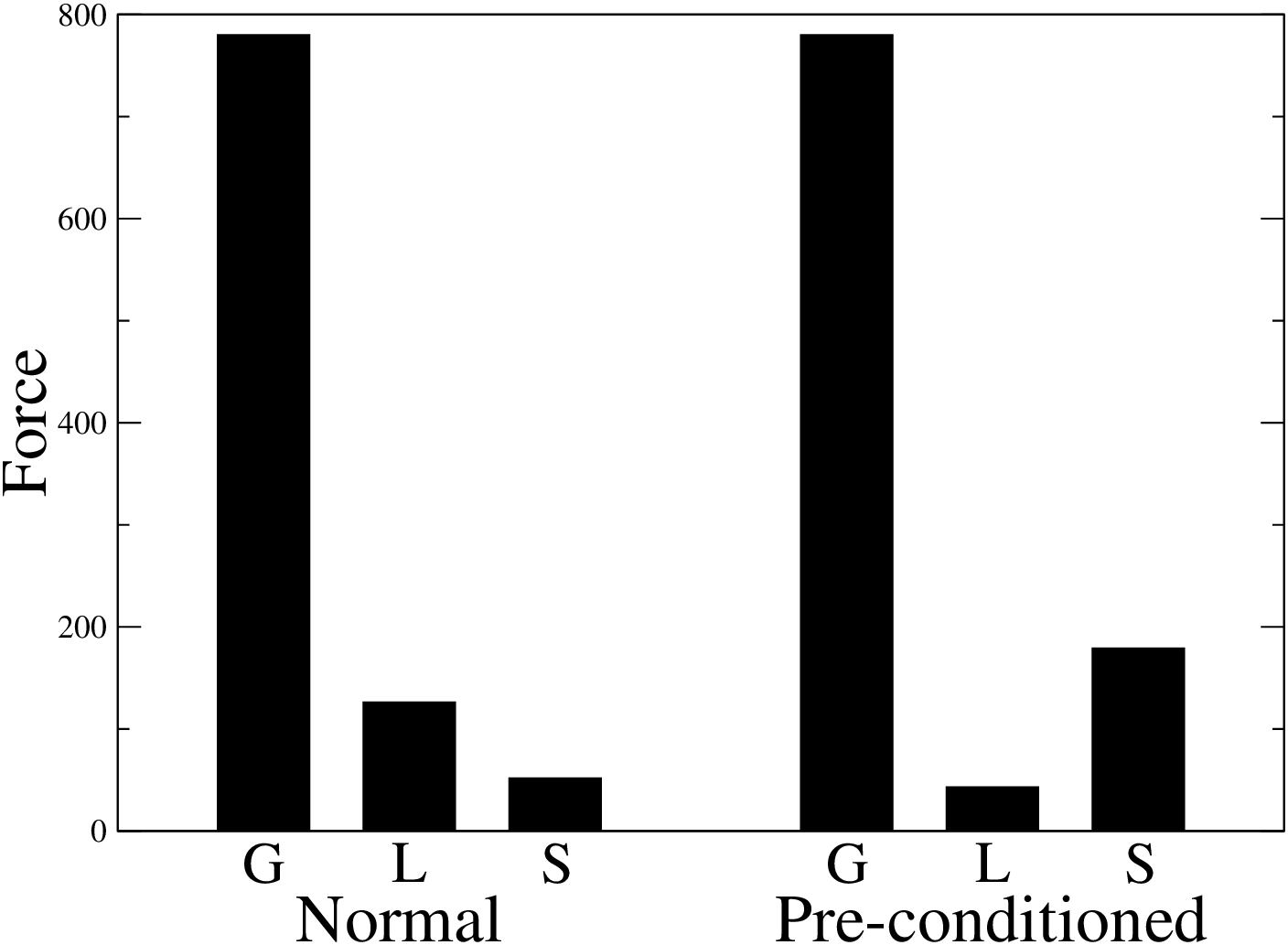}
\label{force}
\vspace{-1.2cm}
\caption{Magnitudes of gauge, light quark, and strange quark forces without quotient pre-conditioning (left) and with quotient pre-conditioning(right)}
\end{minipage}
\vspace{-0.5cm}
\end{figure}

\section{Algorithmic Improvements}
\subsection{Omelyan Integrator}
The Omelyan integrator\cite{Takaishi:2005tz,PhysRevE.66.026701} has much smaller $O(\delta t^2)$ errors compared to the standard leapfrog integrator that is normally used.  As a result, one can use much larger integration step sizes during a molecular dynamics evolution, leading to a 50\% speedup without affecting the acceptance rate. 
\subsection{Multiple time steps}
The gauge field, light quarks, and strange quark contribute different amounts to the force during a molecular dynamics trajectory.  The largest contributions to the force, the gauge force, must be calculated most often, but is actually the least expensive computationally.  By using multiple time scales for the gauge force and the fermion force during integration, we can increase speed while maintaining constant acceptance by tuning the impulses from each force update to approximately the same value.\cite{Sexton:1992nu}.
\subsection{Quotient Pre-conditioning}
The force coming from the light quark and the strange quark are actually quite similar, differing because of small contributions from the lightest modes.  Therefore, we can ``precondition'' the light quark kernel $\mathcal{M}_l$ by dividing by the strange quark kernel $\mathcal{M}_s$\cite{Hasenbusch:2001ne}.
\begin{displaymath}
\mbox{det}(\mathcal{M}_l)^{1/2}\mbox{det}(\mathcal{M}_s)^{1/4} = \left(\frac{\mbox{det}(\mathcal{M}_l)}{\mbox{det}(\mathcal{M}_s)}\right)^{1/2} \mbox{det}(\mathcal{M}_s)^{3/4}
\end{displaymath}
The corresponding action is:
\begin{displaymath}
S_F = \bar{\psi}_l\left(\frac{\mathcal{M}_l}{\mathcal{M}_s}\right)^{1/2}\psi_l + \bar{\psi}_s \mathcal{M}_s^{3/4}\psi
_s
\end{displaymath}
As shown in Fig. 2, quotient preconditioning drastically reduces the force from the light quark, allowing us to calculate the light quark force less often.  Since CG count is dominated by the light quark inversion, this allows for potentially large performance increases.

\begin{table}[t]
\caption{Comparison for different algorithms on finite temperature lattices}
\label{table:1}
\newcommand{\m}{\hphantom{$-$}}
\newcommand{\cc}[1]{\multicolumn{1}{c}{#1}}
\renewcommand{\tabcolsep}{1.0pc} 
\begin{tabular}{lcllrcc}
\hline
\multicolumn{7}{l}{2+1f p4fat3, $m_s a = .065, \beta = 3.30, 16^3\times4, 8^3\times 4$}\\
\hline
& Local & & & $N_{steps}$ & Cost &\\
Algorithm & Volume & Int & $m_q a$ & (fermion) & (kCG/traj) & Acc.\%$^\dagger$\\
\hline
R & 4444 & Leap & .0065 & 192 & 246& -\\
R & 4444 & Leap & .0065 & 50 & 64 & -\\
RHMC & 4444 & Leap & .0065 & 10 & 21.8 & .82\\
RHMC* & 4444 & Omel & .0065 & 5& 12.5 & .70\\
\hline
R & 4444 & Leap & .026 & 50 & 22.5 & -\\
RHMC & 4444 & Leap & .026 & 10 & 6.7& .86\\
RHMC* & 4444 & Omel & .026 & 3 & 3.7& .83\\
\hline
R & 2224 & Leap & .0065 & 192 & 126 & -\\
RHMC & 2224 & Leap & .0065 & 10 & 6.6& .80\\
\hline
R & 2224 & Leap & .026 & 50 & 22.5 & -\\
RHMC & 2224 & Leap & .026 & 8 & 2.5& .80\\
\hline
\multicolumn{7}{l}{*Denotes use of quotient preconditioning}\\
\multicolumn{7}{l}{$\dagger$ A configuration is ``accepted'' with probability = min$(1,\mbox{exp}(-\delta \mathcal{H}))$}
\end{tabular}
\vspace{-0.6cm}
\end{table}

\section{Comparisons of different techniques}
Table \ref{table:1} is a comparison of the different algorithmic techniques.  At the lightest quark mass ($m_q=0.1m_s$), these improvements give a 20x reduction in CG count compared to the R Algorithm.  Even for heavier masses ($m_q=0.4m_s$) we still get a 7x or 8x speedup.

\section{Summary and Future plans}
The use of the RHMC, along with various other algorithmic improvements, have greatly accelerated finite temperature lattice simulations, in some cases gaining as much as a factor of 20 in speed compared to old techniques.  This vastly expands the parameter space accessible to finite temperature lattice simulations with staggered-type fermions.  In the future we hope to test other possible improvements, such as using a $\mathcal{O}(\delta t^4)$ integrator for finite temperature lattices.

\bibliography{sewm06_proceedings}

\section{Acknowledgments}
This work was done on the QCDOC computer with resources provided by Columbia University, Brookhaven National Laboratory, and the RIKEN-Brookhaven Research Center (RBRC).
\end{document}